\documentclass[aps,prb,10pt,twocolumn,amsmath,amssymb,showpacs]{revtex4-1}

\pdfoutput=1    

\usepackage{graphicx}
\usepackage[usenames]{color} 
\usepackage{bm} 
\usepackage{dcolumn}

\begin{document}


\title{Dynamical magnetic charges and linear magnetoelectricity}

\author{Meng Ye}
\email{mengye@physics.rutgers.edu}
\affiliation{Department of Physics and Astronomy, Rutgers University,
Piscataway, New Jersey 08854-8019, USA}

\author{David Vanderbilt}
\affiliation{Department of Physics and Astronomy, Rutgers University,
Piscataway, New Jersey 08854-8019, USA}

\date{January 7, 2014}

\begin{abstract}
Magnetoelectric (ME) materials are of fundamental interest and
have been investigated for their broad potential for technological
applications. The search for, and eventually the theoretical
design of, materials with large ME couplings present challenging
issues. First-principles methods have only recently been
developed to calculate the full ME response tensor $\alpha$
including both electronic and ionic (i.e., lattice-mediated)
contributions. The latter is proportional to both the
Born dynamical electric charge $Z^{\rm e}$ and its analogue, the
dynamical magnetic charge $Z^{\rm m}$. Here we present a
theoretical study of the magnetic charge $Z^{\rm m}$ and the
mechanisms that could enhance it.
Using first-principles density-functional methods, we calculate the
atomic $Z^{\rm m}$ tensors in $\rm{Cr_2O_3}$, a prototypical
magnetoelectric, and in KITP\-ite, a fictitious material that has
previously been reported to show a strong ME response arising from
exchange striction effects.  Our results confirm that in
$\rm{Cr_2O_3}$, the $Z^{\rm m}$ values and resulting ME responses
arise only from spin-orbit coupling (SOC) and are therefore rather
weak.  In KITP\-ite, by contrast, the exchange striction acting on
the non-collinear spin structure induces much $Z^{\rm m}$ values
that persist even when SOC is completely absent.
\end{abstract}

\maketitle


\section{Introduction}

There has been a recent resurgence of research on the magnetoelectric
(ME) effect, which describes the coupling between electricity
and magnetism.\cite{Fiebig} The linear ME effect is defined as
\begin{equation}
\alpha_{\beta\nu}=
\frac{\partial P_\beta}{\partial H_\nu}\Big\vert_{\mathbf{\cal{E}}}
=\mu_0\frac{\partial M_\nu}{\partial{\cal{E}}_\beta}\Big\vert_{\mathbf H}\,,
\end{equation}
where the polarization $\mathbf P$ is linearly induced by an
external magnetic field $\mathbf H$, or the magnetization is
linearly generated by an applied electric field $\mathbf{\cal E}$.
Here indices $\beta$ and $\nu$ denote the Cartesian directions and
$\mu_0$ is the vacuum permeability. This coupling between
electricity and magnetism is of fundamental interest and shows
broad potential for technological applications.

The history of research on the ME effect dates back to the 1960s when
the magnetic symmetry started to be emphasized. It was first
realized by Landau and Lifshitz that the ME response is only
allowed in media without time-reversal symmetry or inversion
symmetry.\cite{Landau} In 1959, Dzyaloshinskii predicted that
$\rm{Cr_2O_3}$ should be a ME crystal\cite{Dzyaloshinskii} based
on its magnetic point group, and experiments successfully measured
the linear induced magnetization by an external electric field
\cite{Astrov,Folen} and the inverse effect.\cite{Rado} The early
theoretical studies and explanations for the ME effect were based
on phenomenological models\cite{phenomen1,phenomen2,phenomen3,
phenomen4} that typically do not distinguish carefully between
microscopic mechanisms.  The recent rapid development of first-principles
methods\cite{Iniguez08,Bousquet11,Malashevich12} has now allowed the
underlying mechanisms in different materials to be classified
and investigated.

The linear ME effect can be decoupled into three contributions, namely
electronic (frozen-ion), ionic (lattice-mediated), and
strain-mediated responses.\cite{Birol12} Each term can be further
subdivided into spin and orbital contributions. The early {\it ab-initio}
studies were focused on the spin-lattice \cite{Iniguez08} and
spin-electronic \cite{Bousquet11} terms. First-principles methods
have only recently been developed to calculate the full ME
response tensor $\alpha$ including both spin and orbital
contributions.\cite{Malashevich12} As the symmetry condition for
the strain-mediated term is more restrictive,
this term is absent in most bulk materials.

Previous studies have shown that the spin-lattice term is dominant
in many materials, as for example in $\rm{Cr_2O_3}$.\cite{Malashevich12}
\'I\~niguez has shown \cite{Iniguez08} that the
lattice contribution can be written as a product of
the Born charge, the force-constant inverse, and the dynamical magnetic
charge, which is the magnetic analogue of the dynamical Born
charge. This dynamical magnetic charge is defined as
\begin{equation}
Z^{\rm m}_{m\nu}=\Omega_0\frac{\partial M_\nu}{\partial
u_m}\Big\vert_{\mathbf{\cal E},\mathbf H,\mathbf\eta} \,.
\end{equation}
Here $\Omega_0$ is the volume of the unit cell containing $N$ atoms
and $u_m$ denotes a periodicity-preserving sublattice displacement,
where $m$
is a composite label running from 1 to $3N$ to represent the atom and
its displacement direction.  The magnetic charge tensor
$Z^{\rm m}$ plays an important role in various lattice-mediated
magnetic responses
and contributes to the Lyddane-Sachs-Teller relationship in
magnetoelectric materials,\cite{Resta-prl11,Resta-prb11}
but the mechanisms that give rise to it are not
yet well understood.  In particular, one route to optimizing the
magnetoelectric coupling is clearly to enhance $Z^{\rm m}$, but
it is not obvious how to do so.

In this work, we use first-principles
density functional methods to study the dynamical magnetic charges in two
materials and explore the different mechanisms responsible for them
in these two cases.
We first study the magnetic charges in $\rm{Cr_2O_3}$, which are
driven by the spin-orbital coupling (SOC) mechanism. Then we study
a fictitious structure, ``KITP\-ite,'' which was reported to have
a large spin-lattice ME coupling according to a previous theory.\cite{KITPite}
The structure of KITP\-ite is such that
the superexchange interactions between
Mn moments are frustrated, leading to a 120$^\circ$ noncollinear
spin structure.
Our study shows that the $Z^{\rm m}$ values, which are orders of magnitude
stronger than in $\rm{Cr_2O_3}$, are responsible for the strong
ME coupling.  We find that this enhancement is present even when SOC
is completely absent, thus confirming that it arises from
exchange striction acting on the non-collinear spins, in contrast to
the case of $\rm{Cr_2O_3}$ where $Z^{\rm m}$ is driven only by
SOC effects.

The paper is organized as follows. In Sec.~\ref{formalism}
we introduce the formalism that describes how the dynamical
magnetic charge tensor enters into the lattice contributions to the
magnetic, ME, and piezomagnetic responses.
In Sec.~\ref{structure&symmetry}, we analyze the structure
and the magnetic symmetry of $\rm{Cr_2O_3}$ and KITP\-ite.
The computational details are described in
Sec.~\ref{first-principles method}. In Sec.~\ref{results},
we present and discuss the computed magnetic charge tensors
for $\rm{Cr_2O_3}$ and KITP\-ite. Finally, Sec.~\ref{summary} provides a
summary.

\section{Preliminaries}

\subsection{Formalism}\label{formalism}

Here, following \'I\~niguez,\cite{Iniguez09} we generalize the
formalism of Wu, Vanderbilt and Hamann\cite{WVH} (WVH) to
include the magnetic field, and use this systematic treatment to
derive the ionic contribution of the ME coupling and other
magnetic properties.

For an insulating system with $N$ atoms in a unit cell, we consider
four kinds of perturbation: (i) a homogeneous electric field ${\bm
{{\cal E}}}$, whose indices $\beta, \gamma$ run over $\{x,y,z\}$;
(ii) a homogeneous magnetic field $\bm H$, whose indices
$\nu,\omega$ also run over $\{x,y,z\}$; (iii) a homogeneous strain
$\bm \eta$, with Voigt indices $i,j=\{1\ldots 6\}$; and
(iv) internal displacements $\bm u$, indexed by composite labels
$m,n$ (atom and displacement direction) running over $1,\ldots,3N$.
In this work we only consider internal displacements that preserve
the bulk periodicity, corresponding to zone-center phonon modes.

The magnetoelectric enthalpy density is defined as
\begin{equation}
E(\bm u, \bm \eta, \bm {{\cal E}}, \bm H) = {\frac1 \Omega_0}
[E^{(0)}_{\rm cell}- \Omega (\bf{\cal E}\cdot \bf{P} +
\mu_0\bf{H}\cdot \bf{M})]\,,
\end{equation}
where $E^{(0)}_{\rm cell}$ is the the zero-field energy per cell
and $\Omega_0$ and $\Omega$ are the undeformed and
deformed cell volumes respectively. $E(\bm u, \bm \eta,
\bm {{\cal E}}, \bm H)$ can be expanded around the zero-field
equilibrium structure as
\begin{equation}
\begin{split}
E=& E_0 + A_m u_m + A_j \eta_j + A_\beta{\cal E}_{\beta}
+ A_\nu H_{\nu}\\
&+ \frac{1}{2} B_{mn}u_m u_n + \frac1 2 B_{jk}\eta_j\eta_k
+ \frac1 2 B_{\beta\gamma}{\cal E}_{\beta}{\cal E}_{\gamma}\\
&+ \frac1 2 B_{\nu\omega}H_{\nu}H_{\omega}
+ B_{mj}u_m\eta_j + B_{m\beta}u_m{\cal E}_\beta \\
&+ B_{m\nu}u_m H_\nu
 + B_{\beta j}{\cal E}_\beta \eta_j + B_{\nu
j }H_\nu \eta_j + B_{\beta\nu}{\cal E}_\beta H_\nu
\end{split}
\label{2nd-order-exp}
\end{equation}
where summation over repeated indices is implied. The
coefficients of the first-order terms correspond to the
atomic forces $F_m=-\Omega_0 A_m$, the
stress tensor $\sigma_j=A_j$, the spontaneous polarization $P^{\rm
S}_\beta=-A_\beta$, and the spontaneous magnetization $M^{\rm
S}_\nu=-\mu_0^{-1} A_\nu$. For the equilibrium structure,
the atomic forces and the stress tensor vanish.
The diagonal second-order coefficients provide
the force-constant matrix $K_{mn}=\Omega_0 B_{mn}$,
the frozen-ion elastic tensor $\bar C_{jk}=B_{jk}$,
the frozen-ion electric susceptibility $\bar\chi^{\rm e}_{\beta\gamma}
  =-\epsilon_0^{-1} B_{\beta\gamma}$,
and the frozen-ion magnetic susceptibility $\bar\chi^{\rm m}_{\nu\omega}
  =-\mu_0^{-1}B_{\nu\omega}$,
where the bar on a quantity indicates a purely electronic response
computed at fixed internal coordinates of the atoms.
The remaining terms correspond to off-diagonal responses, namely
the force-response internal-strain tensor $\Lambda_{mj}=-\Omega_0B_{mj}$,
the frozen-ion piezoelectric tensor $\bar{e}_{\beta j}=-B_{\beta j}$,
the frozen-ion piezomagnetic tensor $\bar{h}_{\nu j }=-\mu_0^{-1}B_{\nu j}$,
the frozen-ion magnetoelectric tensor $\bar\alpha_{\beta\nu}=-B_{\beta\nu}$,
the atomic Born charges
\begin{equation}
Z^{\rm e}_{m\beta}=\Omega_0\frac{\partial P_\beta}{\partial
u_m}\Big\vert_{\mathbf{\cal E},\mathbf H,\mathbf\eta}
=\mu_0^{-1}\frac{\partial F_m}{\partial E_\beta}\Big\vert_{\mathbf{H},\mathbf\eta}
=-\Omega_0B_{m\beta}\,,
\end{equation}
and the atomic magnetic charges
\begin{equation}
Z^{\rm m}_{m\nu}=\Omega_0\frac{\partial M_\nu}{\partial
u_m}\Big\vert_{\mathbf{\cal E},\mathbf H,\mathbf \eta}
=\mu_0^{-1}\frac{\partial F_m}{\partial H_\nu}\Big\vert_{\mathbf{\cal E},\mathbf\eta}
=-\Omega_0\mu_0^{-1}B_{m\nu}\,.
\end{equation}

Static physical responses arise not only from the electronic part
(barred quantities), but also from the ionic contribution associated with
the change of the equilibrium internal displacements $u_m$ with
fields or strain.  The relaxed-ion magnetoelectric enthalpy is
\begin{equation}
\tilde{E}(\bm \eta, \bm {{\cal E}}, \bm H)=\min_{\bm u}E(\bm u,
\bm \eta, \bm {{\cal E}}, \bm H)\,,
\end{equation}
and the minimization is accomplished by substituting
\begin{equation}
u_m=-(B^{-1})_{mn}(B_{nj}\eta_j+B_{n\beta}{\cal E}_\beta+B_{n\nu} H_\nu)
\end{equation}
into Eq.~(\ref{2nd-order-exp}) to obtain the total
relaxed-ion response (including both electronic and ionic parts).
The total relaxed-ion electric susceptibility,
magnetic susceptibility, elastic, piezoelectric, piezomagnetic,
and magnetoelectric tensors are then
\begin{equation}
\chi^{\rm e}_{\beta\gamma}=-\epsilon_0^{-1}\frac{\partial^2
\tilde E}{\partial{\cal E}_\beta \partial{\cal
E}_\gamma}\Big\vert_{\mathbf H, \mathbf \eta}=
\bar\chi^{\rm e}_{\beta\gamma}+\Omega_0^{-1}\epsilon_0^{-1}Z^{\rm
e}_{m\beta}(K^{-1})_{mn}Z^{\rm e}_{n\gamma} \,,
\label{chie}
\end{equation}
\begin{equation}
\chi^{\rm m}_{\nu\omega}=-\mu_0^{-1}\frac{\partial^2 \tilde
E}{\partial H_\nu \partial H_\omega}\Big\vert_{\mathbf{\cal E},
\mathbf \eta}=
\bar\chi^{\rm m}_{\nu\omega}+\Omega_0^{-1}\mu_0 Z^{\rm m}_{m\nu}
(K^{-1})_{mn} Z^{\rm m}_{n\omega} \,,
\label{chim}
\end{equation}
\begin{equation}
C_{jk}=\frac{\partial^2 \tilde E}{\partial\eta_j
\partial\eta_k}\Big\vert_{\mathbf{\cal E}, \mathbf H}=\bar
C_{jk}-\Omega_0^{-1}\Lambda_{mj}(K^{-1})_{mn}\Lambda_{nj} \,,
\end{equation}
\begin{equation}
e_{\beta j}=-\frac{\partial^2 \tilde E}{\partial{\cal
E}_\beta \partial\eta_j}\Big\vert_{\mathbf H}=\bar{e}_{\beta
j}+\Omega_0^{-1} Z^{\rm e}_{m\beta} (K^{-1})_{mn} \Lambda_{nj} \,,
\end{equation}
\begin{equation}
h_{\nu j}=-\frac{\partial^2 \tilde E}{\partial H_\nu
\partial\eta_j}\Big\vert_{\mathbf{\cal E}}=\bar{h}_{\nu j}+
\Omega^{-1}_0 Z^{\rm m}_{m\nu} (K^{-1})_{mn} \Lambda_{nj} \,,
\end{equation}
\begin{equation}
\alpha_{\beta\nu}=-\frac{\partial^2
\tilde E}{\partial{\cal E}_\beta \partial
H_\nu}\Big\vert_{\mathbf\eta}=\bar\alpha_{\beta\nu}+\Omega_0^{-1}\mu_0
Z^{\rm e}_{m\beta} (K^{-1})_{mn} Z^{\rm m}_{n\nu} \,.
\label{alpha}
\end{equation}

\begin{figure}
  \includegraphics[width=8cm]{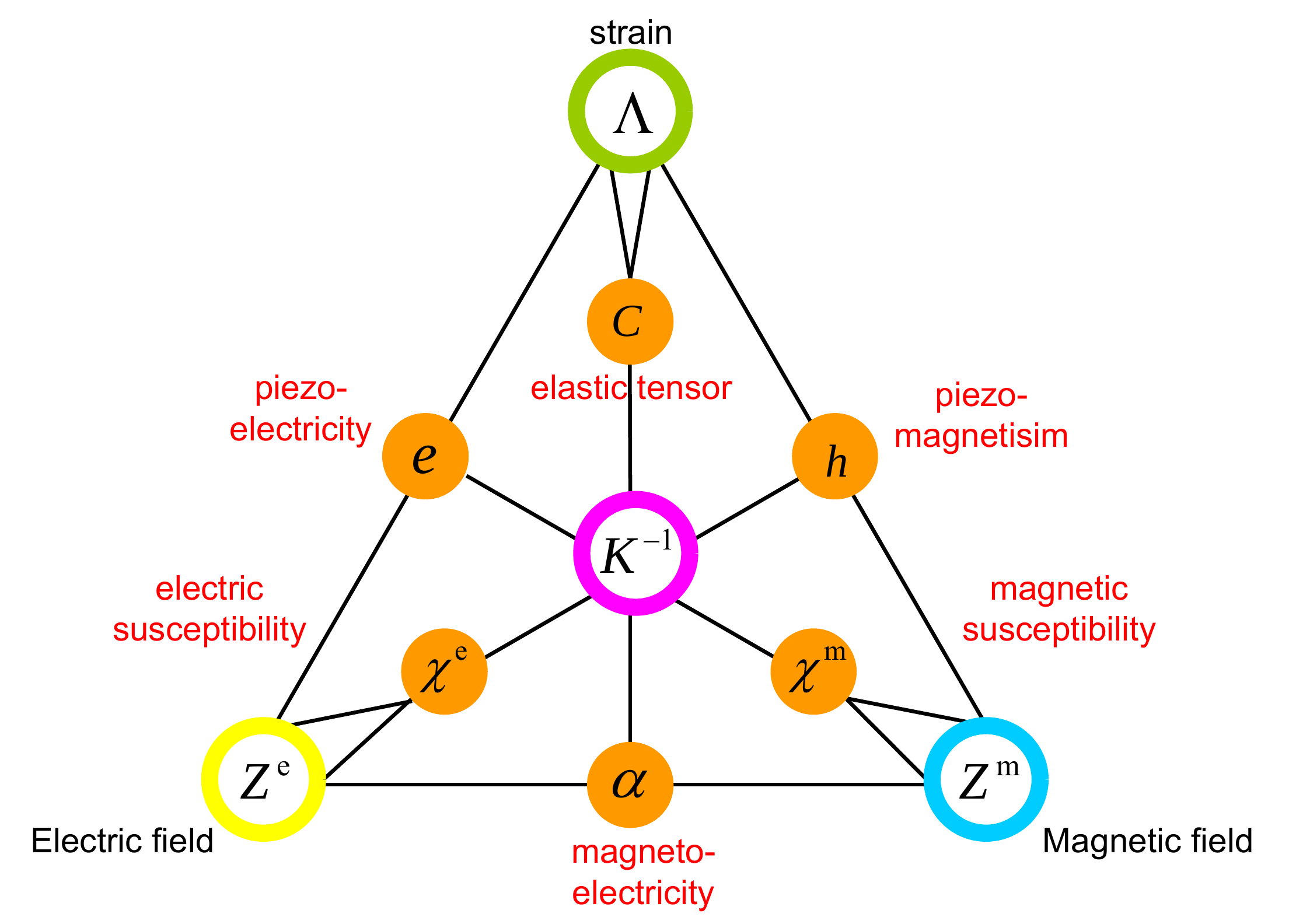}\\
  \caption{(Color online) Sketch showing how the six
  lattice-mediated responses indicated by solid circles (orange)
  are each built up from the four elementary tensors indicated
  by open circles: the Born charge $Z^{\rm e}$ (yellow),
  magnetic charge $Z^{\rm m}$ (blue), internal strain $\Lambda$
  (green), and force-constant inverse $K^{-1}$ (magenta). Each
  lattice-mediated response is given by the product of the three
  elementary tensors connected to it, as indicated explicitly
  in Eqs.~(\ref{chie}-\ref{alpha}).}
  \label{fig:triangle}
\end{figure}

The six lattice-mediated responses in Eqs.~(\ref{chie}-\ref{alpha}) are
all made up of four fundamental tensors: the Born charge tensor $Z^{\rm
e}$, the magnetic charge tensor $Z^{\rm m}$, the internal strain
tensor $\Lambda$, and the inverse force-constant matrix $K^{-1}$.
The manner in which these six lattice responses are
computed from the four fundamental tensors is illustrated in
Fig.~(\ref{fig:triangle}), which depicts the linear-response
connections between elastic, electric and magnetic degrees of
freedom.

If the crystal symmetry is low enough that piezoelectric or
piezomagnetic effects are present, then the strain degrees
of freedom can similarly be eliminated by minimizing the
magnetoelectric enthalpy with respect to them, leading to
additional strain-relaxation contributions to $\chi^{\rm e}$,
$\chi^{\rm m}$, and/or $\alpha$.
  We do not consider these
contributions in the present work because such terms are absent
by symmetry in the materials under consideration here.

The above derivations are carried out in the ($\mathbf{\cal E},\mathbf H$)
frame, which is consistent with the usual experimental conventions.
In the context of first-principles calculations, however, it is more
natural to work in the ($\mathbf{\cal E},\mathbf B$) frame,
as $\mathbf{\cal E}$ and $\mathbf B$ are directly related to
the scalar and vector potentials $\phi$ and $\mathbf A$. The
magnetoelectric tensor
$\alpha$ has different units in these two frames. In the
($\mathbf{\cal E},\mathbf H$) frame, $\alpha$ is defined through
Eq.~(\ref{alpha}) so that the units are s/m. In the
($\mathbf{\cal E},\mathbf B$) frame, $\alpha$ is instead defined as
\begin{equation}
\alpha^{\mathbf{\cal E}\mathbf B}_{\beta\nu}=\frac{\partial M_\nu}
{\partial {\cal E}_\beta}\Big\vert_{\mathbf B}
=\frac{\partial P_\beta}{\partial B_\nu}\Big\vert_{\mathbf{\cal E}}
\end{equation}
and carries units of inverse Ohm,
the same as for $\sqrt{\epsilon_0/\mu_0}$, the
inverse of the impedance of free space. The ME tensors in these
two frames are related by $\alpha^{\mathbf{\cal E}\mathbf H}=
(\mu\alpha)^{\mathbf{\cal E}\mathbf B}$, where $\mu$ is the magnetic
permeability.
The electric and magnetic dynamical charges
in the two frames are related by
$(Z^{\rm e})^{\mathbf{\cal E}\mathbf H}=(Z^{\rm e}+
\alpha\mu Z^{\rm m})^{\mathbf{\cal E}\mathbf B}$
and $(Z^{\rm m})^{\mathbf{\cal E}\mathbf H}=
(\mu Z^{\rm m}/\mu_0)^{\mathbf{\cal E}\mathbf B}$.

For non-ferromagnetic materials we have $\mu\approx\mu_0$, so that the
$Z^{\rm m}$ values are essentially the same in the two frames.
The same is also true for $Z^{\rm e}$, since the product
$(\alpha\mu Z^{\rm m})^{\mathbf{\cal E}\mathbf B}$
is at least five orders of
magnitude smaller than $Z^{\rm e}$
in most magnetoelectric materials.
In this work we report our results in the more conventional
($\mathbf{\cal E},\mathbf H$) frame, even though the
computations are carried out in the ($\mathbf{\cal E},\mathbf B$)
frame.

\subsection{Structure and symmetry}\label{structure&symmetry}

\subsubsection{$\rm{Cr}_2\rm{O}_3$}

\begin{figure}
  \includegraphics[width=6cm]{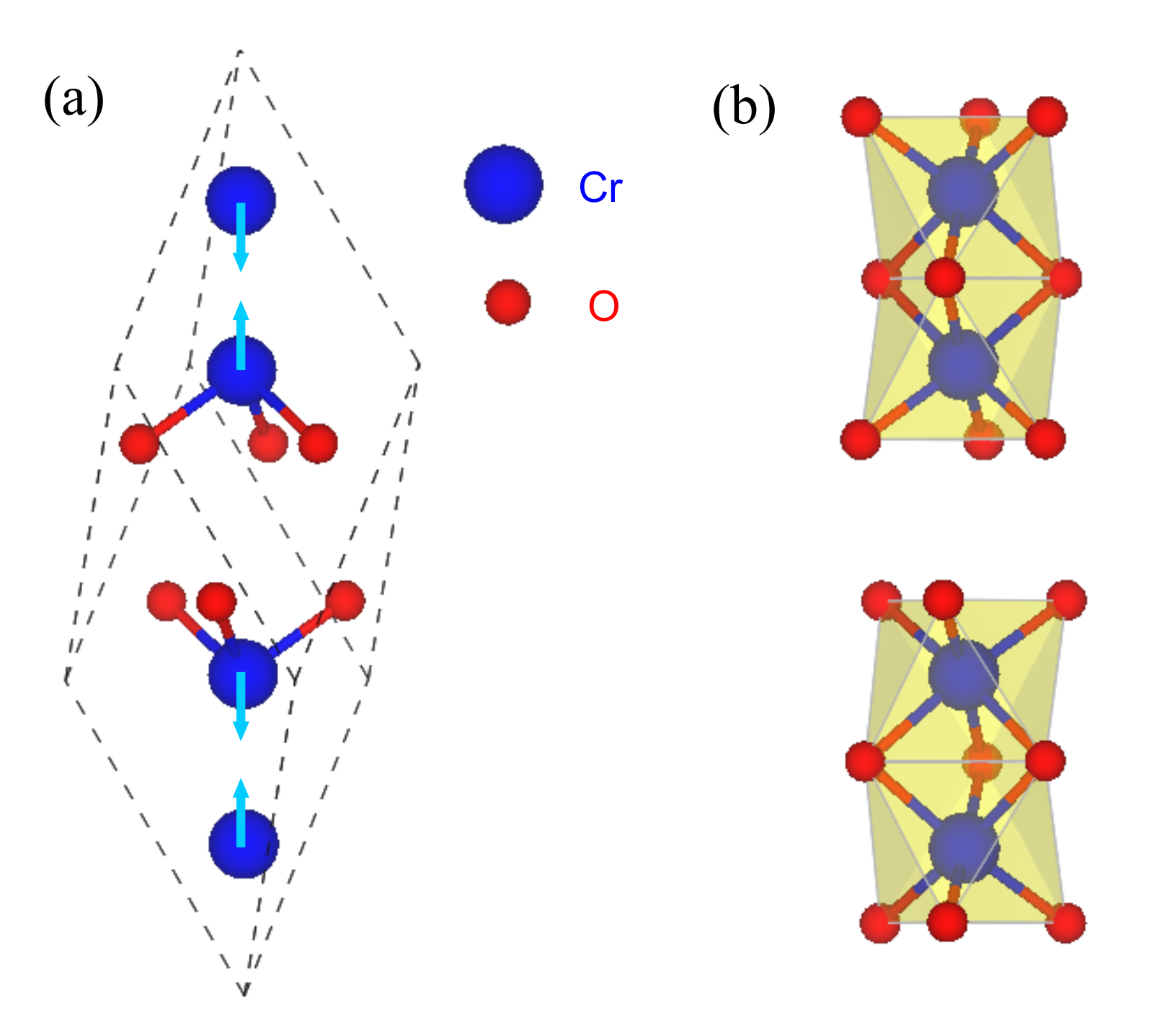}\\
  \caption{(Color online)
  Structure of $\rm{Cr}_2\rm{O}_3$. (a) In the
  rhombohedral primitive cell, four Cr atoms align along the
  the rhombohedral axis with AFM magnetic moments shown by (blue)
  arrows.
  (b) Each Cr atom is at the center of a distorted oxygen octahedron.}
  \label{fig:CrO-Stru}
\end{figure}

\begin{figure}
  \includegraphics[width=8cm]{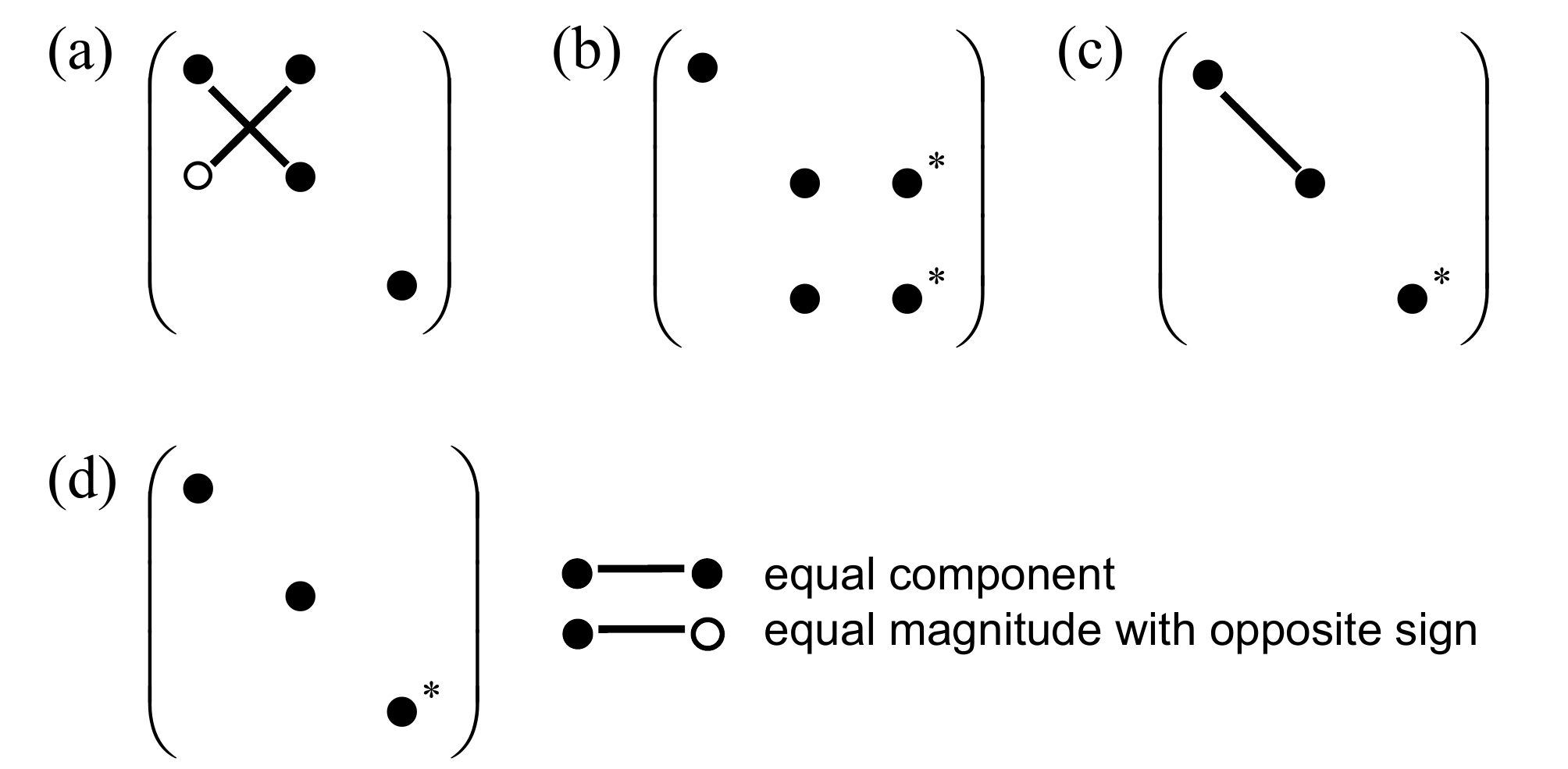}\\
  \caption{Symmetry pattern of the Born and magnetic
  charge tensors for (a) the Cr atom in $\rm{Cr_2O_3}$, (b) the O atom
  in $\rm{Cr_2O_3}$ and the $\rm O^2$ atom in $\rm{CaAlMn_3O_7}$,
  (c) the Ca, Al and $\rm O^1$ atoms in $\rm{CaAlMn_3O_7}$, and
  (d) the Mn and $\rm O^3$ atoms in $\rm{CaAlMn_3O_7}$. The
  elements indicated by an asterisk vanish in the absence of
  SOC for $Z^{\rm m}$ in $\rm{CaAlMn_3O_7}$.}
  \label{fig:tensor-symmetry}
\end{figure}

Cr$_2$O$_3$ adopts the corundum structure
with two formula units per rhombohedral primitive cell as
shown in Fig.~\ref{fig:CrO-Stru}(a). Each Cr atom is at
the center of a distorted oxygen octahedron as shown in
Fig.~\ref{fig:CrO-Stru}(b). It is an antiferromagnetic (AFM)
insulator up to the N\'{e}el temperature $T_{\rm N}=307\,\rm{K}$.
The AFM phase has the magnetic space group
$\rm{R}\bar 3' \rm{c}'$ and the spin
direction on the Cr atoms alternates along the rhombohedral axis.
The magnetic symmetry allows a non-zero ME tensor with two
independent components $\alpha_{\perp}=\alpha_{xx}=\alpha_{yy}$
and $\alpha_{\parallel}=\alpha_{zz}$. Another feature of this
magnetic group is that all the improper rotations are coupled to
the time-reversal operator and vise versa, so that pseudovectors
and ordinary vectors transform in the same way, implying that the magnetic
charge $Z^{\rm m}$ and the Born charge $Z^{\rm e}$ have the
same tensor forms. The three-fold symmetry on each Cr atom restricts
its tensor to have the form shown in
Fig.~\ref{fig:tensor-symmetry}(a).  The symmetry is lower on the O atoms;
for the one  lying on the two-fold rotation axis along $\hat{x}$, for
example, the charge tensor take the form shown in
Fig.~\ref{fig:tensor-symmetry}(b).

\subsubsection{$\rm{KITP\-ite}$}

\begin{figure}
  \includegraphics[width=8cm]{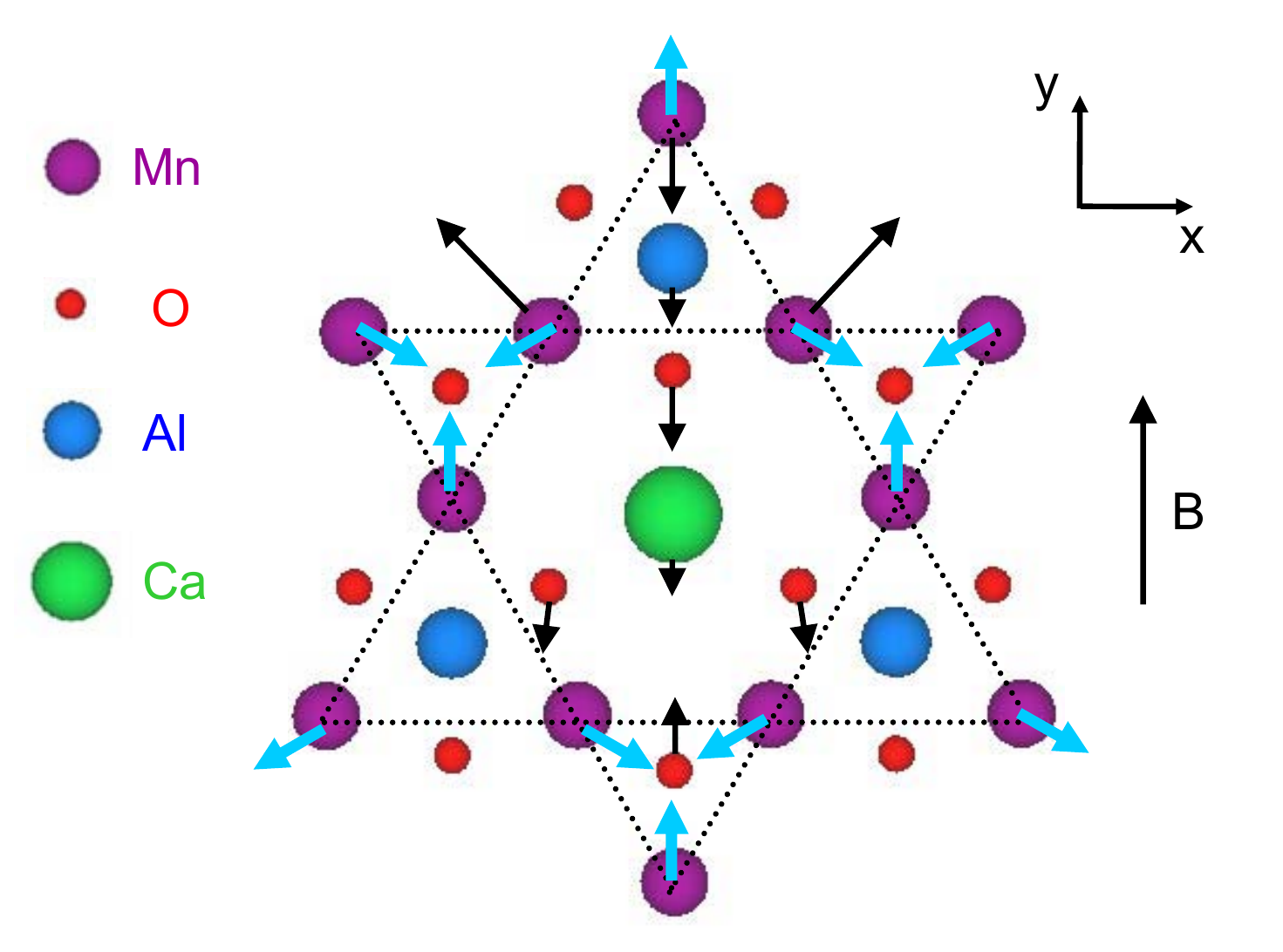}\\
  \caption{(Color online)
  Planar view of the $\rm{CaAlMn_3O_7}$ (KITP\-ite) structure.
  The broad arrows (blue) on the Mn atoms represent the magnetic
  moment directions in the absence of electric or magnetic fields.
  Small (black) arrows indicate the atomic forces induced by an
  external magnetic field applied in the $\hat{y}$ direction.}
  \label{fig:CAMO-force}
\end{figure}

The fictitious ``KITP\-ite'' structure with chemical formula
$\rm{CaAlMn_3O_7}$ is Kagome-like with $120^\circ$ in-plane AFM
spin ordering as showed in Fig.~\ref{fig:CAMO-force}. The unit
cell includes two formula units made by stacking two MnO layers
with $180^\circ$ rotations between layers. Each Mn atom is
surrounded by an oxygen bipyramid and the O atoms are in
three nonequivalent Wyckoff positions: $\rm O^1$ are in the voids
of the Mn triangles; $\rm O^2$ are the apical ions located between
the two MnO layers (not shown in the planar view); and $\rm O^3$ form
the MnO hexagons. The magnetic space group is $\rm{6_3/m'm'c'}$; this has the
same symmetry feature as $\rm{Cr}_2\rm{O}_3$, namely that all the
improper rotations and time-reversal symmetries are coupled
together, so that the Born charges and the magnetic charges follow
the same symmetry restrictions. The charge tensors for Ca, Al and
$\rm O^1$ atoms have the symmetry pattern shown in
Fig.~\ref{fig:tensor-symmetry}(c), and the Mn and $\rm O^3$ atoms
have the charge tensor form of Fig.~\ref{fig:tensor-symmetry}(d).
For the apical $\rm O^2$ atoms, the five independent components in the charge
tensor can be written in the form of Fig.~\ref{fig:tensor-symmetry}(b)
when the on-site two-fold axis is along the $\hat{x}$ direction.

The elements marked by asterisks in Fig.~\ref{fig:tensor-symmetry}
are those that vanish for $Z^{\rm m}$ in $\rm{CaAlMn_3O_7}$ when SOC is
neglected.  The system of magnetic moments is exactly
coplanar in the absence of SOC, and will remain so even after the
application of any first-order nonmagnetic perturbation.
Thus, spin components along $\hat{z}$ cannot be induced, and it
follows that the elements in the third column all vanish in all
atomic $Z^{\rm m}$ tensors in $\rm{CaAlMn_3O_7}$ when SOC is neglected.

\subsection{First-principles methods}\label{first-principles method}

The first-principles calculations for $\rm{Cr_2O_3}$ are performed
with the QUANTUM ESPRESSO \cite{QE} package using the
generalized-gradient approximation parametrized by the
Perdew-Burke-Ernzerhof functional.\cite{PBE}
We employ Troullier-Martin norm-conserving pseudopotentials\cite{TMPP}
with SOC included and Cr $3s$ and $3p$ states incorporated in the
valence.  The wavefunctions are expanded in a
plane-wave basis with cutoff energy 150\,Ry, and a $4\times4\times4$
Monkhorst-Pack k-point mesh is used for the self-consistent-field
loop.
In the Berry-phase polarization calculation,
\cite{ModernPolar} a $4\times4\times5$ k-point sampling is found
to be sufficient. In order to calculate magnetic charges, Born
effective charges and the $\Gamma$-point force-constant matrix,
the finite-difference method is used by displacing each atomic sublattice
in each Cartesian direction and computing the total magnetization,
the Berry-phase polarization and the Hellmann-Feynman forces.
The orbital magnetization is calculated using the modern theory
of orbital magnetization.\cite{ModernOrbMag1,ModernOrbMag2,
ModernOrbMag3}

The calculations for the fictitious KITP\-ite material are
carried out with plane-wave density-functional theory implemented
in VASP.\cite{VASP} PAW pseudopotentials \cite{PAW} with
energy cutoff 400\,eV are sufficient in the non-collinear
magnetization computation without SOC. For the
exchange-correlation functional we use the rotationally
invariant LSDA+U functional,\cite{LDAU} with Hubbard $U=5.5$\,eV
and $J=2.0$\,eV on the $d$ orbitals of the Mn atoms.\cite{UJ} The Born
effective charge tensor and the $\Gamma$-point force-constant matrix
are obtained by linear-response methods. The dynamical magnetic charges are
computed by applying an uniform Zeeman field in the crystal and
computing the resulting forces.  \cite{Bousquet11}
A $4\times4\times4$ Monkhorst-Pack k-point mesh is used in the calculations.

\section{Results}\label{results}

\subsection{$\rm{Cr_2O_3}$}

\begin{table}
\begin{center}
\begin{ruledtabular}
    \caption{Structural parameters of $\rm{Cr_2O_3}$ from
    first-principles calculation and experiments:
    rhombohedral lattice constant $a$, lattice
    angle $\alpha$, and Wyckoff positions for Cr (4c) and O (6e).}
    \begin{tabular}{ l c c c c}
                & & &\multicolumn{2}{c}{Wyckoff position} \\
                & a~({\AA}) & $\alpha$ (deg) & Cr & O     \\ \hline
    This work   & 5.386 & 54.3 & 0.1546 & 0.0617    \\
    Expt.~(Ref.~\onlinecite{CrOstruExp})  & 5.358 & 55.0 & 0.1528 & 0.0566    \\
    \end{tabular}\label{table:CrOstru}
\end{ruledtabular}
\end{center}
\end{table}

\begin{table}[b]
\begin{center}
\begin{ruledtabular}
    \caption{Frequencies (cm$^{-1}$) of zone-center IR-active phonon
    modes of $\rm{Cr_2O_3}$ from first-principles
    calculations and experiments. The two $A_{2u}$ modes are
    longitudinal; the four $E_u$ modes are transverse (doubly
    degenerate).}
    \begin{tabular}{l  c  c  c  c  c  c}
      &\multicolumn{2}{c}{$A_{2u}$ modes} & \multicolumn{4}{c}{$E_u$ modes}\\ \hline
    This work   & 388 & 522 & 297 & 427 & 510 & 610 \\
    Expt.~(Ref.~\onlinecite{chieExp}) & 402 & 533 & 305 & 440 & 538 & 609 \\
    \end{tabular}\label{table:CrOfreq}
\end{ruledtabular}
\end{center}
\end{table}
\begin{table}
\begin{center}
\begin{ruledtabular}
    \caption{Magnetic charges $Z^{\rm{m}}$ ($10^{-2}\mu_{\rm B}/{\rm\AA}$)
    for $\rm{Cr_2O_3}$ in the atomic basis. The magnetic charge
    tensors for Cr and O atoms take the forms shown in
    Figs.~\ref{fig:tensor-symmetry}(a-b).}
    \begin{tabular}{lddccdd}
        &\multicolumn{1}{c}{spin}
        &\multicolumn{1}{c}{orbital} & &
        &\multicolumn{1}{c}{spin}
        &\multicolumn{1}{c}{orbital} \\ \hline
    $Z^{\rm m}_{xx}(\rm{Cr})$ & 5.88 & 0.25 & & $Z^{\rm m}_{yy}(\rm{O})$ &-1.95 &-0.38\\
    $Z^{\rm m}_{xy}(\rm{Cr})$ &-5.69 & 0.02 & & $Z^{\rm m}_{yz}(\rm{O})$ & 0.00 & 1.12\\
    $Z^{\rm m}_{zz}(\rm{Cr})$ & 0.02 & 0.23 & & $Z^{\rm m}_{zy}(\rm{O})$ &-1.10 &-0.72\\
    $Z^{\rm m}_{xx}(\rm{O})$  &-5.92 & 0.06 & & $Z^{\rm m}_{zz}(\rm{O})$ &-0.02 &-0.15\\
    \end{tabular}\label{table:Zm-CrO}
\end{ruledtabular}
\end{center}
\end{table}

The $\rm{Cr}_2\rm{O}_3$ ground-state structural parameters
predicted by our first-principles calculations are in good
agreement with experiment, as shown in Table~\ref{table:CrOstru}.
A group-theory analysis of the long-wavelength phonons shows
that the infrared (IR) active phonon modes, which couple to the
electromagnetic excitations, are the longitudinal $A_{2u}$ modes
and the transverse doubly-degenerated $E_u$ modes,
\begin{equation}
\Gamma_{\rm {IR}}=2A_{2u}+4E_u\,,
\end{equation}
where the acoustic modes have been excluded. The IR-active mode
frequencies shown in Table~\ref{table:CrOfreq} are computed using
linear-response methods, and the results are in good agreement
with experiment.

\begin{table}[b]
\begin{center}
\begin{ruledtabular}
    \caption{Top: Mode decomposition of the Born
    charges $Z^{\rm e}$, and of the spin and orbital contributions
    to the magnetic charges $Z^{\rm m}$, in $\rm{Cr_2O_3}$.
    $C_n$ are the eigenvalues of the force-constant matrix.
    Bottom: Total $A_{2u}$-mode (longitudinal) and $E_u$-mode
    (transverse) elements of the lattice-mediated electric
    susceptibility $\chi^{\rm e}$, magnetic susceptibility
    $\chi^{\rm m}$, and the spin and orbital pars of the ME
    constant $\alpha$.}
    \begin{tabular}{ldddddd}
     &\multicolumn{2}{c}{$A_{2u}$ modes} & \multicolumn{4}{c}{$E_u$ modes}\\ \hline
    $C_n~(\rm{eV}/{\AA}^2)$                                     & 10.5 & 22.9 & 10.2 & 16.0 & 20.2 & 30.9\\
    $Z^{\rm e}~(|\rm{e}|)$                                      & 1.15 & 8.50 & 0.55 & 0.39 & 3.71 & 7.07\\
    $Z^{\rm m}_{\rm spin}~(10^{-2}\mu_{\rm B}/{\rm{\AA}})$       & 0.02 & 0.05 &-0.76 &-3.97 &16.14 &10.55\\
    $Z^{\rm m}_{\rm orb}~(10^{-2}\mu_{\rm B}/{\rm{\AA}})$        & 2.74 &-0.59 & 0.66 &-0.80 &-0.29 & 1.06\\ \hline
    Latt. $\chi^{\rm e}$      &\multicolumn{2}{c}{6.2}  & \multicolumn{4}{c}{4.37} \\
    Latt. $\chi^{\rm m}$      &\multicolumn{2}{c}{$0.05\times10^{-8}$}& \multicolumn{4}{c}{$1.28\times10^{-8}$} \\
    $\alpha_{\rm spin}~(\rm{ps/m})$         &\multicolumn{2}{c}{0.0024} & \multicolumn{4}{c}{0.633} \\
    $\alpha_{\rm orb}~(\rm{ps/m})$          &\multicolumn{2}{c}{0.0097} & \multicolumn{4}{c}{0.025} \\
    \end{tabular}\label{table:CrOresponse}
\end{ruledtabular}
\end{center}
\end{table}

The main results for the magnetic charge tensors of $\rm{Cr_2O_3}$
are reported both in the atomic basis and in the IR-active mode basis in
Tables~\ref{table:Zm-CrO} and \ref{table:CrOresponse}.  The spin
contributions are dominant in the transverse direction, but much
weaker in the longitudinal direction. This is to be expected from
the nearly collinear spin order of $\rm{Cr_2O_3}$, considering that
the magnitudes of the magnetic moments are quite stiff while their
orientations are relatively free to rotate.
The main effect in
the longitudinal direction is from the orbital-magnetization
contribution.  Incidentally,
we also find that the longitudinal components of the magnetic
charge for Cr atoms are very sensitive to the lattice constant
of $\rm{Cr_2O_3}$, especially the Cr-O distance in the
longitudinal direction. Thus, it is essential to choose a proper
exchange-correlation functional to mimic the experimental
ground state structure.

The Born charge tensors for Cr and O are computed to be
$$
Z^{\rm e}({\rm{Cr}}) =
\begin{pmatrix}
3.02 & -0.30 & 0 \\
0.30 &  3.02 & 0 \\
0 & 0 & 3.18
\end{pmatrix}
\rm{e}\,,
$$
$$
Z^{\rm e}(\rm{O}) =
\begin{pmatrix}
-2.36 & 0 & 0 \\
0 & -1.66 & -1.00 \\
0 & -0.88 & -2.12
\end{pmatrix}
\rm{e}\,.
$$
While the symmetry constraints on the non-zero elements
are the same as for $Z^{\rm m}$, the pattern is quite different.
For example, the diagonal elements are of similar magnitude
for $Z^{\rm e}$ but not for $Z^{\rm m}$.

The lattice-mediated magnetic and electric responses for
$\rm{Cr_2O_3}$ computed from  Eqs.~(\ref{chie}-\ref{alpha})
are summarized in the bottom panel of Table~\ref{table:CrOresponse}.
Our computational results are in reasonable agreement with
the experimental room-temperature
lattice-mediated $\chi^{\rm e}_\parallel=4.96$ and $\chi^{\rm
e}_\perp=3.60$ obtained from IR reflectance measurements.\cite{chieExp}
In contrast, the experimentally measured longitudinal and
transverse magnetic susceptibility at low temperature\cite{chimExp}
are on the order of $\sim10^{-3}$,
which is about five orders of magnitude larger than the results
obtained from Eq.~(\ref{chim}).
This difference undoubtedly arises from the fact that the
experimental $\chi^{\rm m}$ is dominated by the electronic
(i.e., frozen-ion) contribution $\bar{\chi}^{\rm m}$ that is
not included in Table~\ref{table:CrOresponse}.
The magnetoelectric response $\alpha_\parallel$
and $\alpha_\perp$ both agree closely
with previous theory, which is in reasonable agreement with
experiment.\cite{Iniguez08,Malashevich12}

\subsection{$\rm{KITPite}$}

\begin{table}
\begin{center}
\begin{ruledtabular}
    \caption{Magnetic charges $Z^{\rm{m}}$ ($10^{-2}\mu_{\rm B}/{\rm\AA}$)
    for $\rm{CaAlMn_3O_7}$ (KITP\-ite) in the atomic basis (spin only).
    The magnetic charge tensors for Ca, Al and $\rm{O^1}$ are of the form
    of Fig.~\ref{fig:tensor-symmetry}(c); those
    for Mn and $\rm{O^3}$ are of the form of Fig.~\ref{fig:tensor-symmetry}
    (d); and that for $\rm{O^2}$ is of the form of
    Fig.~\ref{fig:tensor-symmetry}(b).}
    \begin{tabular}{lddldd}
    \multicolumn{5}{c} {$Z^{\rm{m}}$ ($10^{-2}\mu_{\rm B}/{\rm\AA}$)} \\ \hline
    $Z^{\rm m}_{xx}(\rm{Ca})$ & -43.46  & & $Z^{\rm m}_{xx}(\rm{O^2})$ & -39.15 \\
    $Z^{\rm m}_{xx}(\rm{Al})$ & -24.63  & & $Z^{\rm m}_{yy}(\rm{O^2})$ &   1.23 \\
    $Z^{\rm m}_{xx}(\rm{Mn})$ & 341.53  & & $Z^{\rm m}_{zy}(\rm{O^2})$ & -37.62  \\
    $Z^{\rm m}_{yy}(\rm{Mn})$ &-171.46  & & $Z^{\rm m}_{xx}(\rm{O^3})$ & -56.09 \\
    $Z^{\rm m}_{xx}(\rm{O^1})$ & 66.98  & & $Z^{\rm m}_{yy}(\rm{O^3})$ & -75.23 \\
    \end{tabular}\label{table:Zm-CAMO}
\end{ruledtabular}
\end{center}
\end{table}

When we relax $\rm{KITPite}$ $\rm{CaAlMn_3O_7}$ in the assumed
$\rm{6_3/m'm'c'}$ structure, the unit cell has
a volume of 311.05~$\rm{\AA}^3$ with a $c/a$ ratio of 0.998.
The Wyckoff
coordinates for the Mn atoms (6h) and $\rm{O^3}$ atoms (6g) are
0.5216 and 0.1871. Other atoms are in high-symmetry Wyckoff
positions. The IR-active modes are
\begin{equation}
\Gamma_{\rm {IR}}=6A_{2u}+9E_{1u}
\end{equation}
excluding the acoustic modes. The longitudinal
$A_{2u}$ modes do not contribute to the magnetic response when
spin-orbit interaction is absent in $\rm{CaAlMn_3O_7}$, because
the longitudinal components of the magnetic charges $Z^{\rm m}$
are zero.

\begin{table}
\begin{center}
\begin{ruledtabular}
    \caption{The Born charges $Z^{\rm e}$ and the magnetic
    charges $Z^{\rm m}$ for the IR-active $A_{2u}$ modes in
    $\rm{CaAlMn_3O_7}$.
    $C_n$ are the eigenvalues of the force-constant matrix.}
    \begin{tabular}{ddd}
  \multicolumn{1}{c}{$C_n~(\rm{eV}/{\rm\AA}^2)$} &
  \multicolumn{1}{c}{$Z^{\rm e}~(|\rm e|)$} &
  \multicolumn{1}{c}{$Z^{\rm m}_{\rm spin}~(10^{-2}\mu_{\rm B}/{\rm\AA})$} \\
  \hline
-2.4 & 3.7 & 539.7 \\
-1.1 & 4.7 &  17.2 \\
 2.8 & 4.3 &  -0.6 \\
 7.1 & 2.4 & 266.4 \\
11.6 & 5.1 &-107.8 \\
12.0 & 2.4 & -74.5 \\
35.3 & 7.9 & -15.9 \\
46.7 & 2.2 &  34.8 \\
55.1 & 4.6 &-325.7 \\
    \end{tabular}\label{table:CAMOresponse}
\end{ruledtabular}
\end{center}
\end{table}

The results for the magnetic charge tensors are reported in
the atomic basis and the IR-active mode basis in
Tables~\ref{table:Zm-CAMO} and \ref{table:CAMOresponse}
respectively. The calculated force-constant eigenvalues and
Born charges $Z^{\rm e}$ are also listed in Table~\ref{table:CAMOresponse}.
The Born charges in KITP\-ite and $\rm{Cr_2O_3}$ are all close to the
atomic valence charge values. As the KITP\-ite structure is
fictitious and two $E_{1u}$ modes are unstable in the
high-symmetry structure, we will focus on the results for the
magnetic charges and omit any discussion of the the magnetic and
dielectric responses.

The magnetic charges in the KITP\-ite structure are found to be
much larger than for $\rm{Cr}_2\rm{O}_3$. For the transition-metal
ion, the magnetic charge of Mn in KITP\-ite is $\sim$50 times
larger than for Cr in $\rm{Cr}_2\rm{O}_3$.
The magnetic charges in $\rm{Cr}_2\rm{O}_3$ are driven by
SOC, which acts as an antisymmetric exchange
field. Thus, the weakness of the SOC on the Cr atoms
implies that the magnetic charges and magnetic responses
are small in $\rm{Cr}_2\rm{O}_3$. In the KITP\-ite structure,
we deliberately exclude spin-orbit interaction, so the magnetic
charges are purely induced by the spin frustration and the
super-exchange between Mn-O-Mn atoms. This exchange striction
mechanism causes the
magnetic charges in $\rm{CaAlMn_3O_7}$ to be dozens of times larger
than the SOC-driven responses in $\rm{Cr}_2\rm{O}_3$.

Since the orbital magnetization is strongly quenched in
most $3d$ transition metals, we expect the orbital contribution
to the $Z^{\rm m}$ tensors in $\rm{CaAlMn_3O_7}$ to be
comparable with those in $\rm{Cr}_2\rm{O}_3$, i.e.,
on the order of $10^{-2}\,\mu_{\rm B}/{\rm\AA}$.
Since this is $\sim$2 orders of magnitude smaller than
the typical spin contribution in $\rm{CaAlMn_3O_7}$, we have not
included it in our calculation.  The main point of our
study of KITP\-ite $\rm{CaAlMn_3O_7}$ has been to demonstrate
that exchange-striction effects can give rise to large
$Z^{\rm m}$ values based on a mechanism that does not
involve SOC at all.

\section{Summary}\label{summary}

In summary, we have begun by presenting a systematic formulation
of the role played by the dynamic
magnetic charge tensor $Z^{\rm m}$ in the lattice magnetic,
magnetoelectric, and piezomagnetic responses of crystalline solids.
We have then used first-principles density-functional methods
to compute the atomic $Z^{\rm m}$ tensors for two prototypical
materials, namely $\rm{Cr_2O_3}$, a well-studied magnetoelectric
material, and fictitious KITP\-ite, which displays a very large
lattice ME effect.  We find that the physics is quite different in
the two cases, with mechanisms based on SOC giving only small
$Z^{\rm m}$ values in the collinear antiferromagnet $\rm{Cr_2O_3}$,
while exchange-striction effects induce very large $Z^{\rm m}$'s
in noncollinear KITP\-ite.

Our calculations are part of a broader effort to identify
mechanisms that could induce large magnetic charge values.
They help to reinforce a picture in which
SOC effects give only weak contributions, at least in $3d$
transition-metal compounds, whereas exchange striction
can induce much larger effects in materials in which spin
frustration gives rise to a noncollinear spin structure.
In this respect, the conclusions parallel those that have
emerged with respect to the polarization in multiferroics and
magnetically-induced improper ferroelectrics, where exchange
striction, when present, typically produce much larger
effects than SOC.\cite{Cheong-Mostovoy}

Our work points to some possible future directions for exploration.
One obvious direction is to identify experimentally known
materials in which exchange striction gives rise to
large $Z^{\rm m}$ values.  In such systems, lattice-mediated
effects might even contribute significantly to the magnetic
susceptibility; while such contributions are normally neglected
for $\chi^{\rm m}$, we note that $Z^{\rm m}$ appears to the
second power in Eq.~(\ref{chim}), so this contribution might
be significant, especially in soft-mode systems.
It might also be interesting to explore the role of
these magnetic charges in the phenomenology of electromagnons.\cite{Pimenov}
Finally, we point out that, unlike $Z^{\rm e}$, $Z^{\rm m}$ remains
well-defined even in metals; while magnetoelectric
effects do not exist in this case, it would still be interesting
to explore the consequences of large $Z^{\rm m}$ values
in such systems.

\acknowledgments
This work was supported by ONR Grant N00014-12-1-1035.



\begin{thebibliography}{}
\bibitem{Fiebig} M. Fiebig, J. Phys. D {\bf 38}, R123 (2005).

\bibitem{Landau} L. D. Landau and E. M. Lifshitz, {\it Electrodynamics of continuous media} (1957)
(English Transl., Pergamon Press, Oxford, 1960).

\bibitem{Dzyaloshinskii} I. E. Dzyaloshinskii, Sov. Phys. JETP {\bf 10}, 628 (1959).

\bibitem{Astrov} D. N. Astrov, Sov. Phys. JETP {\bf 11}, 708 (1960).

\bibitem{Folen} V. J. Folen, G. T. Rado and E. W. Stalder, Phys. Rev. Lett. {\bf 6}, 607 (1961).

\bibitem{Rado} G. T. Rado and V. J. Folen, Phys. Rev. Lett. {\bf 7}, 310 (1961).

\bibitem{phenomen1} G. T. Rado, Phys. Rev. Lett. {\bf 6}, 609 (1961).

\bibitem{phenomen2} G. T. Rado, Phys. Rev. {\bf 128}, 2546 (1962).

\bibitem{phenomen3} R. Hornreich and S. Shtrikman, Phys. Rev. {\bf 161}, 506 (1967).

\bibitem{phenomen4} O. F. de Alcantara Bonfim and G. A. Gehring, Adv. Phys. {\bf 29}, 731 (1980).

\bibitem{Iniguez08} J. \'{I}\~{n}iguez, Phys. Rev. Lett. {\bf 101}, 117201 (2008).

\bibitem{Bousquet11} E. Bousquet, N. A. Spaldin, and K. T. Delaney, Phys. Rev. Lett. {\bf 106}, 107202 (2011).

\bibitem{Malashevich12} A. Malashevich, S. Coh, I. Souza and D. Vanderbilt, Phys. Rev. B {\bf 86}, 094430 (2012).

\bibitem{Birol12} T. Birol, N. A. Benedek, H. Das, A. L. Wysocki, A. T. Mulder, B. M. Abbett, E. H. Smith, S. Ghosh, and C. J. Fennie, Curr. Opin. Solid State Mater. Sci. {\bf 16}, 227 (2012).

\bibitem{Resta-prl11} R. Resta, Phys. Rev. Lett. {\bf 106}, 047202 (2011).

\bibitem{Resta-prb11} R. Resta, Phys. Rev. B {\bf 84}, 214428 (2011).

\bibitem{KITPite} K. T. Delaney, M. Mostovoy and N. A. Spaldin, Phys. Rev. Lett {\bf 102}, 157203 (2009).

\bibitem{WVH} X. Wu, D. Vanderbilt and D. R. Hamann, Phys. Rev. B {\bf 72}, 035105 (2005).

\bibitem{Iniguez09} J. C. Wojdel and J. \'{I}\~{n}iguez, Phys. Rev. Lett. {\bf 103}, 267205 (2009).

\bibitem{QE} P. Giannozzi {\it et al.}, J. Phys.: Condens. Matter {\bf 21}, 395502 (2009).

\bibitem{PBE} J. P. Perdew, K. Burke, and M. Ernzerhof, Phys. Rev. Lett. {\bf 77}, 3865 (1996).

\bibitem{TMPP} N. Troullier and J. L. Martins, Phys. Rev. B {\bf 43}, 1993 (1991).

\bibitem{ModernPolar} R. D. King-Smith and D. Vanderbilt, Phys. Rev. B {\bf 47}, 1651 (1993).

\bibitem{ModernOrbMag1} D. Xiao, J. Shi, and Q. Niu, Phys. Rev. Lett. {\bf 95}, 137204 (2005).

\bibitem{ModernOrbMag2} T. Thonhauser, D. Ceresoli, D. Vanderbilt, and R. Resta, Phys. Rev. Lett. {\bf 95}, 137205 (2005).

\bibitem{ModernOrbMag3} D. Ceresoli, T. Thonhauser, D. Vanderbilt, and R. Resta, Phys. Rev. B {\bf 74}, 024408 (2006).

\bibitem{VASP} G. Kresse and J. Furthm\"{u}ller, Phys. Rev. B {\bf 54}, 11169 (1996).

\bibitem{PAW} P. E. Blochl, Phys. Rev. B {\bf 50}, 17953(1994); G. Kresse and D. Joubert, Phys. Rev. B {\bf 59}, 1758 (1999).

\bibitem{LDAU} A. I. Liechtenstein, V. I. Anisimov, and J. Zaanen, Phys. Rev. B {\bf 52}, R5467 (1995)

\bibitem{UJ} Z. Yang, Z. Huang, L. Ye, and X. Xie, Phys. Rev. B {\bf 60}, 15674 (1999).

\bibitem{chieExp} G. Lucovsky, R. J. Sladek, and J. W. Allen, Phys. Rev. B {\bf 16}, 4716 (1977).

\bibitem{chimExp} S. Foner, Phys. Rev. {\bf 130}, 183 (1963)

\bibitem{CrOstruExp} A. H. Hill, A. Harrison, C. Dickinson, W. Zhou, and W. Kockelmann, Microporous Mesoporous Mater. {\bf 130}, 280 (2010).

\bibitem{Cheong-Mostovoy} S.-W. Cheong and M. Mostovoy, Nature
Mater. {\bf 6}, 13 (2007).

\bibitem{Pimenov} A. Pimenov, A. M. Shuvaev, A. A. Mukhin, and A. Loidl,
J. Phys. Cond. Matt. {\bf 20}, 434209 (2008).

\end{thebibliography}
\end{document}